\documentstyle[aps,preprint,floats,tighten]{revtex}
\input{psfig.tex}
\begin{document}
\draft
\preprint{\vbox{\hbox{CU-TP-767} 
                \hbox{CAL-615}
                \hbox{astro-ph/9609132}
}}

\title{A Probe of Primordial Gravity Waves and Vorticity}

\author{Marc Kamionkowski\footnote{kamion@phys.columbia.edu}}
\address{Department of Physics, Columbia University, 538 West
120th Street,
New York, New York~~10027}
\author{Arthur Kosowsky\footnote{akosowsky@cfa.harvard.edu}}
\address{Harvard-Smithsonian Center for Astrophysics,
60 Garden Street, Cambridge, Massachusetts~~02138
\\and\\
Department of Physics, Lyman Laboratory, Harvard University,
Cambridge, Massachusetts~~02138}
\author{Albert Stebbins\footnote{stebbins@perseus.fnal.gov}}
\address{NASA/Fermilab Astrophysics Center, Fermi National
Accelerator Laboratory, Batavia, Illinois 60510-0500}
\date{September 1996}
\maketitle

\begin{abstract}
A formalism for describing an all-sky map of the polarization of the cosmic
microwave background is presented.  The polarization pattern on the sky
can be decomposed into two geometrically distinct components. One of these
components is not coupled to density inhomogeneities.  A  non-zero amplitude
for this component of polarization can only be caused by tensor or
vector metric perturbations.
This allows unambiguous identification of long-wavelength
gravity waves or large-scale vortical flows at the time of last scattering.
\end{abstract}

\pacs{98.70.V, 98.80.C}

\def\hatn{{\bf \hat n}}
\def\hatnprime{{\bf \hat n'}}
\def\hatnone{{\bf \hat n_1}}
\def\hatntwo{{\bf \hat n_2}}
\def\vecx{{\bf x}}
\def\veck{{\bf k}}
\def\hatx{{\bf \hat x}}
\def\hatk{{\bf \hat k}}
\def\VEV#1{{\langle #1 \rangle}}
\def\cP{{\cal P}}
\long\def\comment#1{}

With the COBE detection of large-angle anisotropy in the cosmic microwave
background (CMB), results from numerous balloon-borne and ground-based
experiments, and the advent of a new generation of satellite missions, the CMB
is becoming an increasingly precise probe of the early Universe.  CMB
anisotropies will help determine whether density perturbations (scalar modes)
are the result of inflation, topological defects, or perhaps some other
mechanism.  Detection of a stochastic gravity-wave background (tensor modes)
\cite{abbott} or vortical motions in the primeval fluid (vector modes) would
help to discriminate between these models.  Inflation damps out vector modes
but will produce some tensor modes and also predicts a specific relationship
between the spectrum of the scalar and tensor fluctuations\cite{steinhardt}.
In contrast, topological defects will produce a mixture of scalar, vector, and
tensor modes. Scalar modes give rise to both the observed large-scale structure
and CMB fluctuations, while in the foreseeable future we can only expect to
observe the consequences of tensor or vector modes through their effects on the
CMB.  Without a model of primordial fluctuations, the contribution of scalar,
vector, and tensor modes to the CMB  temperature anisotropy are
indistinguishable.

However, any mechanism which produces temperature
anisotropies will invariably lead to non-zero polarization as
well\cite{compton,tensors,arthur}. 
As we demonstrate in this Letter, this polarization signal can be
used to discriminate between scalar and vector or tensor
metric perturbations. COBE has already mapped the polarization
pattern with an angular resolution of $7^\circ$ (although the data has not been
analyzed), and MAP \cite{MAP} will measure the polarization with a resolution
of around $0.3^\circ$.

In prior work, the auto- and cross-correlations between the Stokes parameters
$Q$ and $U$ and the temperature $T$ have been considered.  Here $Q$ and $U$ are
defined with respect to particular 
orthogonal axes on the celestial sphere.  While this
formalism does provide a complete description of the polarization, there
is no rotationally invariant way to lay down orthogonal basis vectors on a
sphere, so the meaning of $QQ$, $QU$, $UU$, $QT$, or $UT$ will 
depend on absolute
positions of the points being correlated rather than just the relative
position.  Calculations of this sort have been done with a small-angle
approximation, since rotational non-invariance disappears 
when considering only
a small patch of the sky.  However, this formalism is not optimal
for describing the complete temperature
and polarization correlations present in full-sky maps.

Here we present a rotationally covariant formalism
for describing the polarization pattern on a full sky.  The Stokes parameters,
defined by the $2\times2$ correlation matrix of the electric field of incoming
photons, can be described as a tensor field on the celestial sphere.  The $Q$
and $U$ parameters, describing linear polarization, are just given by the 
symmetric trace-free (STF) part of this tensor.  For example, in spherical
polar coordinates $(\theta,\phi)$, where the spherical metric is 
$g_{ab}={\rm diag}(1,\sin^2\theta)$, the polarization tensor is
\begin{equation}
  {\cal P}_{ab}(\hatn)={1\over 2} \left( \begin{array}{cc}
   \vphantom{1\over 2}Q(\hatn) & -U(\hatn) \sin\theta \\
   -U(\hatn)\sin\theta & -Q(\hatn)\sin^2\theta \\
   \end{array} \right).
\label{whatPis}
\end{equation}
The ``$ab$'' are the tensor indices, and we use standard tensor notation
throughout. It is natural to decompose the linear-polarization
pattern into STF tensor spherical
harmonics\cite{zerilli,albert}, which constitute a complete orthonormal set of
rank-2 STF tensors on the sphere.  There are two types of harmonic
STF tensors, $Y_{(lm)ab}^{{\rm G}}$ and $Y_{(lm)ab}^{{\rm C}}$, one of each for
every one of the usual spherical harmonics $Y_{(lm)}$ with $l\ge 2$.
Two sets of  tensor harmonics are
required as there are two modes of linear polarization, $Q$ and $U$.
Since Compton scattering can produce no net circular polarization, the
CMB is expected to have $V=0$, and the $V$ Stokes parameter will not
be considered further.

The harmonic expansion of an all-sky map of the CMB temperature and
polarization can be written
\begin{eqnarray}
 {T(\hatn) \over T_0} & = & 1 + \sum_{lm} a^{\rm T}_{(lm)} Y_{(lm)}(\hatn), \cr
 {{\cal P}_{ab}(\hatn)\over T_0} & = &
 \sum_{lm} \left[ a_{(lm)}^{{\rm G}}Y_{(lm)ab}^{{\rm G}}(\hatn)
                 +a_{(lm)}^{{\rm C}}Y_{(lm)ab}^{{\rm C}}(\hatn) \right].
\label{Expansion}
\end{eqnarray}
The mode amplitudes are given by
\begin{eqnarray}
a^{\rm T}_{(lm)}&=&{1\over T_0}\int d\hatn\,T(\hatn)\,Y_{(lm)}^*(\hatn),\cr
a^{\rm G}_{(lm)}&=&{1\over T_0}\int d\hatn\,{\cal P}_{ab}(\hatn)\,
                                         Y_{(lm)}^{{\rm G} \,ab\, *}(\hatn),\cr
a^{\rm C}_{(lm)}&=&{1\over T_0}\int d\hatn\,{\cal P}_{ab}(\hatn)\,
                                          Y_{(lm)}^{{\rm C} \, ab\, *}(\hatn),
\label{Amplitudes}
\end{eqnarray}
which can be derived from the orthonormality properties
\begin{eqnarray}
\int d\hatn\,Y_{(l m )}^*(\hatn)
             Y_{(l'm')}  (\hatn)&=&\delta_{ll'}\delta_{mm'},                \cr
\int d\hatn\,Y_{(lm)ab}^{{\rm G}\,*}(\hatn)
             Y_{(l'm')}^{{\rm G}\,\,ab}(\hatn)&=&\delta_{ll'}\delta_{mm'},  \cr
\int d\hatn\,Y_{(lm)ab}^{{\rm C}\,*}(\hatn)
             Y_{(l'm')}^{{\rm C}\,\,ab}(\hatn)&=&\delta_{ll'}\delta_{mm'},  \cr
\int d\hatn\,Y_{(lm)ab}^{{\rm G}\,*}(\hatn)
             Y_{(l'm')}^{{\rm C}\,\,ab}(\hatn)&=&0.
\label{Orthonormality}
\end{eqnarray}
Here $T_0$ is the
cosmological mean CMB temperature and we are assuming $Q$ and
$U$ are measured in brightness temperature units rather than
flux units.

The two geometrically distinct tensor harmonics are
\begin{eqnarray}
Y_{(lm)ab}^{\rm G}&=&N_l
                     \left(Y_{(lm):ab}-{1\over2}g_{ab}Y_{(lm):c}{}^c\right),\cr
Y_{(lm)ab}^{\rm C}&=&{N_l\over 2}
                 \left(\vphantom{1\over 2}Y_{(lm):ac}\epsilon^c{}_b 
                       +                  Y_{(lm):bc}\epsilon^c{}_a \right).
\label{TensorHarmonics}
\end{eqnarray}
Here $N_l=\sqrt{{2(l-2)!/(l+2)!}}$ is a normalization factor, $\epsilon_{ab}$
is the completely antisymmetric tensor, and ``:'' indicates a covariant
derivative on the sphere.  In two dimensions, any STF tensor can be
uniquely decomposed into a
part of the form $A_{:ab}-(1/2)g_{ab} A_{:c}{}^c$ and another part of the form
$B_{:ac}\epsilon^c{}_b+B_{:bc}\epsilon^c{}_a$ where $A$ and $B$ are two scalar
functions.   This decomposition is quite similar to the
decomposition of a 
vector field into a part which is the gradient of a scalar field 
and a part which is
the curl of a vector field; hence we use the notation
G for ``gradient'' and C for
``curl''.  Since the $Y_{(lm)}$'s provide a complete basis for scalar
functions on the sphere, the $Y_{(lm)ab}^{\rm G}$'s and
$Y_{(lm)ab}^{\rm C}$'s provide a complete basis for G-type and C-type STF
tensors,
respectively.  This G/C decomposition is also known as the scalar/pseudo-scalar
decomposition \cite{albert}.

In $(\theta,\phi)$ coordinates, where Eq.~(\ref{whatPis}) holds, the harmonics
are given explicitly by
\begin{eqnarray}
     Y_{(lm)ab}^{\rm G}({\bf\hat n})&=&{N_l\over 2}
     \left(
     \begin{array}{cc} W_{(lm)} & X_{(lm)}\sin\theta\\
                       X_{(lm)}\sin\theta & -W_{(lm)}\sin^2\theta\\
     \end{array}
     \right), \cr
\phantom{:}\cr
     Y_{(lm)ab}^{\rm C}({\bf\hat n})&=&{N_l\over 2}
     \left(
     \begin{array}{cc} -X_{(lm)} & W_{(lm)}\sin\theta\\
                       W_{(lm)}\sin\theta & X_{(lm)}\sin^2\theta
       \end{array}
       \right)
\label{YGCexplicit}
\end{eqnarray}
with the definitions
\begin{eqnarray}
     W_{(lm)}&\equiv&\left( {\partial^2 \over \partial\theta^2} -
     \cot\theta {\partial \over \partial\theta} +
     {m^2\over\sin^2\theta}\right)Y_{(lm)},                 \cr
\phantom{:}\cr
     X_{(lm)}&\equiv&{2im\over \sin\theta}
     \left( {\partial \over \partial\theta} -
     \cot\theta \right) Y_{(lm)}.
\label{XWdefn}
\end{eqnarray}
The exchange symmetry $\{Q,U\}\leftrightarrow\{U,-Q\}$ as
G$\leftrightarrow$C indicates that $Y_{(lm)ab}^{\rm G}$ and
$Y_{(lm)ab}^{\rm C}$ represent polarizations rotated by
$45^\circ$.

A most useful property of the G/C decomposition is that, in linear theory,
scalar perturbations can produce only G-type polarization and not C-type
polarization.  This is in contrast to tensor or vector metric perturbations
which will produce a mixture of both types.  To understand why scalar metric
perturbations do not produce a C-type polarization pattern, consider a scalar
perturbation with single Fourier mode $\veck$ in the ${\bf \hat z}$ direction.
The polarization in a given direction can be represented by a magnitude 
${\cal P}=(Q^2 + U^2)^{1/2}$ and an orientation angle $\alpha$ from the axis
defining the Stokes parameters (here, choose $\hat\theta$), where 
$\tan2\alpha=U/Q$. 
For scalar perturbations, the orientation of the polarization can
be determined only by the direction of $\bf k$: thus $\alpha=0$ if the
polarization orientation is along the direction of $\bf k$, or $\alpha=\pi/2$
if the orientation is perpendicular to the direction of $\bf k$. In either
case, in a given region of the sky all of the orientations are parallel and
thus the polarization pattern has no curl. Since the curl is a linear operator,
summing over Fourier modes does not alter this conclusion. For tensor and
vector perturbations, the azimuthal symmetry in the scalar case is explicitly
broken, and thus the Fourier vector does not completely define the direction of
the polarization orientation. Another way to state this argument is that scalar
perturbations have no handedness so they cannot produce any ``curl'', whereas
vector and tensor perturbations {\it do} have a handedness and therefore can.

	Finding a non-zero component of C-type polarization in the CMB would
provide compelling evidence for significant contribution of either vector or
tensor perturbations at the time of last scattering.  Given a polarization map
of even a small part of the sky one could in principle test for vector or
tensor contribution by computing the combination of derivatives of the
polarization given by ${\cal P}^{ab}{}_{:bc}\epsilon^c{}_a$ which will be
non-zero only for C-type polarization, i.e. when vector or tensor perturbations
are present.  Similarly only G-type polarization 
contributes to ${\cal P}^{ab}{}_{:ab}$.  
Of course, taking derivatives of noisy data is
problematic; more robust measures are given below.

We now turn to statistics of CMB polarization.  If the cosmological
inhomogeneities are Gaussian random noise, then to the extent linear theory 
is valid, the CMB fluctuations will also be Gaussian random noise. 
Regardless of whether the distribution is Gaussian,
rotational invariance requires that the 2-point correlations be of the form
\begin{eqnarray}
\left\langle a_{(lm)}^{\rm T\,*}\,a_{(l'm')}^{\rm T}\right\rangle
                                     &=&C_l^{\rm T}\delta_{ll'}\delta_{mm'},\cr
\left\langle a_{(lm)}^{\rm G\,*}\,a_{(l'm')}^{\rm G}\right\rangle
                                     &=&C_l^{\rm G}\delta_{ll'}\delta_{mm'},\cr
\left\langle a_{(lm)}^{\rm C\,*}\,a_{(l'm')}^{\rm C} \right\rangle
                                     &=&C_l^{\rm C}\delta_{ll'}\delta_{mm'},\cr
\left\langle a_{(lm)}^{\rm T\,*}\,a_{(l'm')}^{\rm G}\right\rangle
                                    &=&C_l^{\rm TG}\delta_{ll'}\delta_{mm'},\cr
\left\langle a_{(lm)}^{\rm T\,*}\,a_{(l'm')}^{\rm C}\right\rangle
                                    &=&C_l^{\rm TC}\delta_{ll'}\delta_{mm'},\cr
\left\langle a_{(lm)}^{\rm G\,*}\, a_{(l'm')}^{\rm C}\right\rangle
                                    &=&C_l^{\rm GC}\delta_{ll'}\delta_{mm'}\ .
\label{TwoPoint}
\end{eqnarray}
If we also require that the distribution of inhomogeneities be invariant under
parity, then $C_l^{\rm TC}=C_l^{\rm GC}=0$ 
since the $Y_{(lm)}$ and the $Y_{(lm)ab}^{\rm G}$ have parity
$(-1)^l$ while the $Y_{(lm)ab}^{\rm C}$ have parity $(-1)^{l+1}$.  
Measuring a non-zero $C_l^{\rm TC}$ and/or $C_l^{\rm GC}$
would be quite interesting, 
indicating a handedness to the inhomogeneities in our universe. However we
do not expect this and will henceforth only consider the four angular power
spectra $\{C_l^{\rm T},C_l^{\rm G},C_l^{\rm C},C_l^{\rm TG}\}$.
The first is the
well-known angular power spectrum of 
temperature anisotropies while the last three, new to
this paper, are related to various quantities in previous work (see
Ref.~\cite{inpreparation}).  Note that the scalar, vector, and
tensor contribution to each of the $C_l$'s adds in quadrature, i.e. for $X=$T,
G, C, TG 
\begin{equation}
C_l^X=C_l^{X\rm scalar}+C_l^{X\rm vector}+C_l^{X\rm tensor}\ ,
\label{qudrature}
\end{equation}
and this is true whether or not the fluctuations are Gaussian.  We have argued
that $C_l^{\rm C\,scalar}=0$.

Given an all-sky temperature/polarization map, one can determine the
$a_{(lm)}$'s using Eq.~(\ref{Amplitudes}), and then construct estimators for
the $C_l$'s in the usual way, i.e.,
\begin{eqnarray}
\widehat{C_l^{\rm T }}=\sum_{m=-l}^l{|a_{(lm)}^{\rm T}|^2\over2l+1}, &\qquad&
\widehat{C_l^{\rm G }}=\sum_{m=-l}^l{|a_{(lm)}^{\rm G}|^2\over2l+1},        \cr
\widehat{C_l^{\rm C }}=\sum_{m=-l}^l{|a_{(lm)}^{\rm C}|^2\over2l+1}, &\qquad&
\widehat{C_l^{\rm TG}}=\sum_{m=-l}^l{ a_{(lm)}^{\rm T\,*}\,
                                      a_{(lm)}^{\rm G}   \over2l+1}.
\label{ClPesimators}
\end{eqnarray}
If only part of the sky is mapped, the same techniques developed
to analyze anisotropy with incomplete sky coverage \cite{incomplete} may be
applied to polarization to construct other estimators of the various $C_l$'s.
The mean square polarization is
\begin{equation}
\overline{Q^2+U^2}=2\overline{{\cal P}^{ab} {\cal P}_{ab}}
={\cal P}_{\rm G}^2+{\cal P}_{\rm C}^2
\label{meansquareP}
\end{equation}
where
\begin{equation}
{\overline{{\cal P}_{\rm G}^2}\over T_0^2}
                =\sum_{l=2}^\infty {2l+1\over8\pi} \widehat{C_l^{\rm G}},\qquad
{\overline{{\cal P}_{\rm C}^2}\over T_0^2}
                =\sum_{l=2}^\infty {2l+1\over8\pi} \widehat{C_l^{\rm C}} .
\label{mostpowerful}
\end{equation}
Since scalar modes do not contribute to $\overline{{\cal P}_{\rm C}^2}$, this
statistic provides a powerful and unambiguous model-independent
probe of tensor and vector perturbations. 

To test a given spectrum of tensor modes against a polarization map,
comparing the complete set of predicted $C_l^{{\rm C}}$ with the
estimators $\widehat{C_l^{\rm C}}$ is more powerful than considering
only ${\cal P}_{\rm C}^2$, if the detection has sufficient signal-to-noise.  
Usually, however, the theory being tested has scalar as well as non-scalar
modes, along with undetermined
cosmological parameters.  If so, the most information can be
extracted from the map by comparing the entire set of predicted moments, 
$\{C_l^{\rm T},C_l^{\rm G},C_l^{\rm C},C_l^{\rm TG}\}$, with the measured
estimators \cite{kamspergelsug,parameters}.

Note that only the $C_l^{\rm C}$'s potentially allow detection 
of a small vector or
tensor signal.  If scalar perturbations dominate, then the vector or tensor
signal in $\{C_l^{\rm T},C_l^{\rm G},C_l^{\rm TG}\}$ may be swamped by the
cosmic variance in the scalar signal, but the $C_l^{\rm C}$'s are not
contaminated in this way.  The cross-correlation
moments $C_l^{\rm TG}$, which differ for scalar, vector, and
tensor perturbations \cite{crosssign}, will be larger than the
polarization autocorrelation moments.  Furthermore, the noise in
the temperature and polarization maps will be to a large extent
uncorrelated, so the pixel-noise variance to these
moments will be substantially reduced.  Therefore, the
temperature-polarization cross correlation may be measured with
some precision.

Much of the small-angle formalism of
Refs. \cite{compton,tensors,arthur,crosssign,cross,ng,uros} can
be reproduced by
replacing the $Y_{(lm)}(\hatn)$'s in our formalism with Fourier modes,
$e^{i\bf{l}\cdot\hatn}$, and using regular derivatives rather than 
covariant ones.   This small-angle formalism is completely analogous to that
developed above and will provide an accurate description of a region of
sky small enough to be approximated by a flat surface.  The G/C decomposition 
in the small-angle formalism can be used to detect non-scalar perturbations on
small scales, though the tensor and vector signal are liable to drop off
rapidly at angular scales smaller than a few degrees.

To make contact with previous work, we can write the two-point temperature
and polarization correlation functions
\cite{compton,tensors,arthur,crosssign,cross,ng,uros} in terms of multipole
moments\cite{inpreparation}.  Although correlation functions of Stokes
parameters which appear in the previous literature depend on the positions
of the points being correlated, rotationally-invariant
correlation functions exist
which are closely related to those discussed above. To
construct them, define Stokes parameters $Q_r$ and $U_r$ with respect
to axes which are
parallel and perpendicular to the great arc (or geodesic) which connects the
two points being correlated.  The two-point correlation functions are
\begin{eqnarray}
C^T (\theta) &=& \VEV{T(\hatn_1)
                      T(\hatn_2)}_{\hatn_1\cdot\hatn_2=\cos\theta}          \cr
      =&T_0^2&\sum_l{2l+1\over4\pi}\,C_l^{\rm T}\,P_l(\cos\theta),          \cr
C^Q (\theta) &=& \VEV{Q_r(\hatn_1)
                      Q_r(\hatn_2)}_{\hatn_1\cdot\hatn_2=\cos\theta}        \cr
      =&T_0^2&\sum_l{2l+1\over4\pi}[ C_l^{\rm G} W_{(l2)}(\theta,0)
                                    +iC_l^{\rm C} X_{(l2)}(\theta,0)],      \cr
C^U (\theta) &=& \VEV{U_r(\hatn_1)
                      U_r(\hatn_2)}_{\hatn_1\cdot\hatn_2=\cos\theta}        \cr
      =&T_0^2&\sum_l{2l+1\over4\pi}[ C_l^{\rm C} W_{(l2)}(\theta,0)
                                     +iC_l^{\rm G} X_{(l2)}(\theta,0)],     \cr
C^{TQ}(\theta) &=& \VEV{T(\hatn_1)
                      Q_r(\hatn_2)}_{\hatn_1\cdot\hatn_2=\cos\theta}        \cr
      =&T_0^2&\sum_l\,{2l+1\over4\pi}\,N_l\,C_l^{\rm TG}\,P_l^2(\cos\theta),
\label{Correlations}
\end{eqnarray}
where $\theta$ is the angle separating the two points and $P_l^2$ is the
$m=2$
associated Legendre function.  Since $Q_r$ and $T$ are invariant under
reflection along the great arc connecting the two points while $U_r$
changes sign,
$\VEV{Q_rU_r}=\VEV{U_rT}=0$ if statistical invariance under
parity holds.  Eqs.~(\ref{Correlations}) reduce to the correct small-angle
formulae \cite{uros} when $\theta\ll1$.  The functions
$\{C^T(\theta),C^Q(\theta),C^U(\theta),C^{TQ}(\theta)\}$ are a different way of
representing $\{C_l^{\rm G},C_l^{\rm T},C_l^{\rm C},C_l^{\rm TG}\}$ and
vice versa.  In Gaussian models either set provides a complete statistical
description of the temperature and polarization patterns.

The largest hurdles to detecting and characterizing CMB polarization are
sensitivity and foregrounds.  In adiabatic models with standard recombination
the polarization is only a few percent of the anisotropy, although it may be
larger in reionized \cite{ng}, isocurvature, or topological-defect models.
Thus the polarization signal is at least an order of magnitude below current
experimental sensitivities. Experiments planned or envisioned over the coming
decade, however, will likely attain the raw sensitivity necessary for detailed
polarization investigations. Polarized emission from foreground sources is a
relatively unknown factor at this time. Foreground emission and any Faraday
rotation will certainly contribute to the C-type polarization, but these
contaminants can be subtracted using multi-frequency observations.  On 
sub-degree scales, where the signal from vector and tensor modes are liable to
be negligible, any measurable C-type polarization is a likely indicator of
contamination. Polarization measurements will be difficult, but the promise of
using them to detect gravity waves or vorticity, and hence to discriminate
between cosmological models, makes these measurement potentially 
very valuable.

\smallskip

We thank G. Jungman and U. Seljak for helpful discussions.
This work was supported by D.O.E. contract DEFG02-92-ER 40699, NASA NAG5-3091,
and the Alfred P. Sloan Foundation at Columbia, NASA AST94-19400 at FNAL, and
the Harvard Society of Fellows.  M.K. acknowledges the hospitality of the
NASA/Fermilab Astrophysics Center and the CERN Theory Group.  A.S. acknowledges
the hospitality of the Aspen Center for Physics and the support of the NASA
grant NAG 5-2788.

\end{document}